\documentclass[a4paper,aps,prd,10pt,preprintnumbers,showpacs,twocolumn,superscriptaddress,nofootinbib,amsmath,amssymb]{revtex4-1}
\usepackage[utf8]{inputenc}
\usepackage[T1]{fontenc}
\usepackage{cmap}

\def\m{\overline{m}}
\def\a{\widetilde{\alpha}}

\begin{document}
\title{$4D$ Einstein-Lovelock black holes: Hierarchy of orders in curvature}

\author{R. A. Konoplya}\email{roman.konoplya@gmail.com}
\affiliation{Institute of Physics and Research Centre of Theoretical Physics and Astrophysics, Faculty of Philosophy and Science, Silesian University in Opava, Bezručovo nám. 13, CZ-746 01 Opava, Czech Republic}
\affiliation{Peoples Friendship University of Russia (RUDN University), 6 Miklukho-Maklaya Street, Moscow 117198, Russian Federation}

\author{A. Zhidenko}\email{olexandr.zhydenko@ufabc.edu.br}
\affiliation{Institute of Physics and Research Centre of Theoretical Physics and Astrophysics, Faculty of Philosophy and Science, Silesian University in Opava, Bezručovo nám. 13, CZ-746 01 Opava, Czech Republic}
\affiliation{Centro de Matemática, Computação e Cognição (CMCC), Universidade Federal do ABC (UFABC),\\ Rua Abolição, CEP: 09210-180, Santo André, SP, Brazil}

\begin{abstract}
The Einstein-Lovelock theory contains an infinite series of corrections to the Einstein term with an increasing power of the curvature.
It is well-known that for large black holes the lowest (Gauss-Bonnet) term is the dominant one, while for smaller black holes higher curvature corrections become important. We will show that if one is limited by positive values of the coupling constants, then the dynamical instability of black holes serves as an effective cut-off of influence of higher curvature corrections in the 4D Einstein-Lovelock approach: the higher is the order of the Lovelock term, the smaller is the maximal value of the coupling constant allowing for stability, so that effectively only a first few orders can deform the observable values seemingly. For negative values of coupling constants this is not so, and, despite some suppression of higher order terms also occurs due to the decreasing threshold values of the coupling constant, this does not lead to an noticeable opportunity to neglect higher order corrections. In the case a lot of orders of Lovelock theory are taken into account, so that the black-hole solution depends on a great number of coupling constants, we propose a compact description of it in terms of only two or three parameters encoding all the observable values.
\end{abstract}
\maketitle

\section{Introduction}

According to the Lovelock theorem only metric and Einstein tensors are divergence free, symmetric, and concomitant of the metric tensor and its derivatives in four dimensions \cite{Lovelock:1971yv,Lovelock:1972vz}. Therefore, it was concluded that the appropriate vacuum equations in $D=4$ are the Einstein equations. In $D>4$ the theory of gravity is generalized by adding higher curvature corrections \cite{Lovelock:1971yv} to the Einstein term. Motivated by the low-energy limit of string theory and higher dimensional gravity, black hole in the $D>4$ Einstein-Gauss-Bonnet gravity and its Lovelock generalization \cite{Lovelock:1971yv} were extensively studied and a number of interesting features were observed. The life-time of even a softly Gauss-Bonnet corrected black hole proved out to be much longer due to a strong suppression of Hawking radiation \cite{Konoplya:2010vz}.
Another remarkable feature is that small (relatively the coupling constants at the higher curvature terms) Einstein-Lovelock black holes are unstable \cite{Dotti:2005sq,Gleiser:2005ra,Konoplya:2008ix}. Counter-intuitively the instability develops at high multipole numbers, while the lowest multipoles remain stable. Near the event horizon the corresponding effective potential has a negative gap, which becomes deeper for large multipole numbers so that larger multipoles are more unstable \cite{Takahashi:2011qda,Takahashi:2012np,Yoshida:2015vua}. Such a behavior is a consequence of nonhyperbolicity of the perturbation equations in the instability region \cite{Reall:2014pwa}, manifesting itself as the absence of convergence of the sum over multipole numbers. Since the instability corresponds to the regime of long wavelengthes, it was called \emph{the eikonal instability} \cite{Cuyubamba:2016cug,Konoplya:2017lhs,Konoplya:2017zwo}. It is also a simple example when gravitational perturbations break down the correspondence between the eikonal quasinormal modes and characteristics of null geodesics \cite{Cardoso:2008bp,Konoplya:2017wot}.

Recently, there has been suggested the way to bypass the Lovelock's theorem \cite{Glavan:2019inb} by performing a kind of dimensional regularization of the Gauss-Bonnet equations and obtain an effectively four-dimensional metric theory of gravity with diffeomorphism invariance and second order equations of motion. The theory is formulated in $D > 4$ dimensions and then, the four-dimensional effective theory is defined as the limit $D \to 4$ of the higher-dimensional theory after the re-scaling of the coupling constant $\alpha \to \alpha/(D-4)$. It is interesting to note that, prior to \cite{Glavan:2019inb}, the dimensional regularization of the Einstein-Gauss-Bonnet theory was suggested in~\cite{Tomozawa:2011gp}.

This approach was generalized to the $4D$ Einstein-Lovelock gravity in \cite{Konoplya:2020qqh,Casalino:2020kbt}. Various properties of black holes in this context, such as (in)stability, quasinormal modes and shadows, were considered for the first time in \cite{Konoplya:2020bxa}, while the innermost circular orbits were analyzed in \cite{Guo:2020zmf}. The generalization to the charged black holes and an asymptotically anti-de Sitter and de Sitter cases in the $4D$ Einstein-Gauss-Bonnet theory was considered in \cite{Fernandes:2020rpa}. Some further properties of black holes for this novel theory, such as axial symmetry, Hawking radiation and thermodynamics, linear perturbations, stability, collapse, vacuum solutions and others were considered in \cite{Wei:2020ght,Kumar:2020owy,Hegde:2020xlv,Zhang:2020qew,Lu:2020iav,Konoplya:2020ibi,Ghosh:2020syx,Konoplya:2020juj,Kobayashi:2020wqy,Zhang:2020qam,Kumar:2020uyz,HosseiniMansoori:2020yfj,Wei:2020poh,Singh:2020nwo,Churilova:2020aca,Mishra:2020gce,Heydari-Fard:2020sib,Konoplya:2020cbv,Jin:2020emq,Zhang:2020sjh,EslamPanah:2020hoj,NaveenaKumara:2020rmi,Aragon:2020qdc,Malafarina:2020pvl,Yang:2020czk,Fernandes:2020nbq,Cuyubamba:2020moe,Mahapatra:2020rds,Shu:2020cjw,Casalino:2020pyv,Liu:2020evp,Devi:2020uac,Ma:2020ufk,Liu:2020yhu,Kumar:2020sag,Churilova:2020mif,Ge:2020tid,Zeng:2020dco}.

Let us emphasise that the approach suggested in \cite{Glavan:2019inb} is essentially a regularization scheme. It is formulated in $D>4$ dimensional spacetime and the $D \to 4$ limit gives nothing, but the second order differential equations for the metric tensor which does not guarantee that the various tensor identities which we used to see in $4D$ General Relativity, for example, the Bianchi identity, will have the same form. Therefore, when one considers the matter which is propagating in the background of the $4D$ metric, the obtained $4D$ limit for the metric is valid, while when the matter is non-minimally coupled to gravity, like it happens, for example, for gravitational perturbations and analysis of stability, then $D$-dimensional perturbation equations must be the starting point, as it was done, for example, in \cite{Konoplya:2020bxa,Konoplya:2020juj}. This, in our opinion rather evident fact, was emphasised in \cite{Gurses:2020ofy}. Another evident observation is that not every higher-dimensional solution allow for the four-dimensional regularization, simply because there may be no four dimensional analogue of the corresponding higher dimensional system. For example, in order to obtain the rotating black hole solution in $4D$, one cannot consider a higher-dimensional black hole with multiple momenta in different directions, but rather the higher dimensional black hole with a single momentum on the brane. A qualitatively similar illustration was suggested in \cite{Tian:2020nzb}. None of the above observations disproves the dimensional regularization suggested in \cite{Glavan:2019inb}. Recently in \cite{Arrechea:2020evj} it was claimed that non-linear perturbations cannot be regularized in the same way, although the linear perturbations do not have this problem \cite{Konoplya:2020bxa,Konoplya:2020juj}. Apparently non-linear perturbations and the initial value problem must be further studied within the above approach. The facts that the four-dimensional black hole metric obtained when searching for quantum corrections to the entropy \cite{Cai:2009ua,Cognola} was reproduced via dimensional regularization scheme in \cite{Glavan:2019inb} and the BTZ-like black brane found in \cite{Konoplya:2020ibi} was recently reproduced via adding extra degrees of freedom (a scalar field) in the well-defined four dimensional scalar-tensor theory \cite{Hennigar:2020} with the Gauss-Bonnet term apparently signify that the above regularization can be an effective tool at least in some cases and further study of the limits of its applicability is appealing.
It has been recently shown within the Bondi-Sachs framework that the corresponding scalar-tensor Hordenski gravity has no scalar propagator in low-dimensional theories \cite{Lu:2020mjp}, indicating thereby that the stability analysis performed in \cite{Konoplya:2020juj,Cuyubamba:2020moe} is also valid for such theories.

An essential question arises when dealing with the Einstein-Gauss-Bonnet theory as the low-energy limit of string theory: How good approximation the Gauss-Bonnet correction is, that is, can we ignore the infinite series by neglecting all the higher orders of the Lovelock expansion except the lowest Gauss-Bonnet one? As the coupling constant of $m$-th order comes with the denominator $r_{H}^{-2m+2}$, where $m=2$ corresponds to the Gauss-Bonnet term, then for sufficiently small black holes the immediate answer is no!
At the same time, for sufficiently large black holes even the Gauss-Bonnet term must be a good approximation.  Apparently, between these two extreme regimes a number of Lovelock terms must be important and our aim here is to understand further features of black holes when a number of higher curvature corrections with various coupling constants are added. Using the previously announced by us general analysis of the region of eikonal (in)stability \cite{Konoplya:2020juj} we first show that if one is limited by positive values of coupling constants in front of Lovelock terms, then the eikonal instability works as an effective cut-off for higher order terms, because the larger is the order of the Lovelock correction, the smaller is the critical value of the coupling constant for the onset of instability. We show that basic observable quantities such as quasinormal modes, radius of the black-hole shadow, frequencies at the innermost stable circular orbit etc. change almost indistinguishably when the Lovelock corrections of higher than the fourth order in curvature are included.

This effective cut-off due to the instability does not take place for negative coupling constants, for which case, however, the influence of higher orders in curvature are also suppressed, though at a much smaller rate, so that corrections of even the 10th and higher orders in curvature can still be distinguished from the Schwarzschild limit. At this point we come to another problem: How to study properties of a black hole metric in the Einstein-Lovelock theory with a lot of Lovelock terms, that is, containing many coupling constants as parameters of the metric? The effective way to solve this problem has been recently suggested in \cite{Konoplya:2020hyk}, where, using the generic parametrization \cite{Rezzolla:2014mua,Konoplya:2016jvv} it was shown that astrophysically relevant black holes whose observable quantities in the radiation zone can be distinguished from those in the Einstein theory can usually be very well approximated by only a few parameters. Here we apply this approach for analysis of $4D$ Einstein-Lovelock black holes with a lot of Lovelock terms.

The paper is organized as follows. Sec.~\ref{sec:blackhole} gives brief information on $4D$ Einstein-Lovelock regularization and the black-hole metric under consideration. Sec.~\ref{sec:cutoff} discusses the effective cut-off of higher curvature terms owing to the eikonal instability and influence of Lovelock terms at higher orders upon observable quantities. Sec.~\ref{sec:param} suggests a compact description of the black hole metric which depends upon a lot of coupling constants in terms of only a few parameters. Finally, in the Conclusions we summarize the obtained results and mention some open questions.

\section{Static black holes in the four-dimensional Lovelock theory}\label{sec:blackhole}
The Lagrangian density of the Einstein-Lovelock theory has the form
\cite{Lovelock:1971yv}:
\begin{eqnarray}\label{Lagrangian}
  \mathcal{L} &=& -2\Lambda+\sum_{m=1}^{\m}\frac{1}{2^m}\frac{\alpha_m}{m}
  \delta^{\mu_1\nu_1\mu_2\nu_2 \ldots\mu_m\nu_m}_{\lambda_1\sigma_1\lambda_2\sigma_2\ldots\lambda_m\sigma_m}\,\\\nonumber
  &&\times R_{\mu_1\nu_1}^{\phantom{\mu_1\nu_1}\lambda_1\sigma_1} R_{\mu_2\nu_2}^{\phantom{\mu_2\nu_2}\lambda_2\sigma_2} \ldots R_{\mu_m\nu_m}^{\phantom{\mu_m\nu_m}\lambda_m\sigma_m},
\end{eqnarray}
where $\delta^{\mu_1\mu_2\ldots\mu_p}_{\nu_1\nu_2\ldots\nu_p}$
is the generalized totally antisymmetric Kronecker delta, $R_{\mu\nu}^{\phantom{{\mu\nu}}\lambda\sigma}$ is the Riemann tensor, $\alpha_1=1/8\pi G=1$ and $\alpha_2,\alpha_3,\alpha_4,\ldots$ are arbitrary constants of the theory.

The Euler-Lagrange equations, corresponding to the Lagrangian density (\ref{Lagrangian}) read \cite{Kofinas:2007ns}:
\begin{eqnarray}\label{Lovelock}\nonumber
  \Lambda\delta^{\mu}_{\nu} &=& R^{\mu}_{\nu}-\frac{R}{2}\delta^{\mu}_{\nu}+\sum_{m=2}^{\m}\frac{1}{2^{m+1}}\frac{\alpha_m}{m}
  \delta^{\mu\mu_1\nu_1\mu_2\nu_2 \ldots\mu_m\nu_m}_{\nu\lambda_1\sigma_1\lambda_2\sigma_2\ldots\lambda_m\sigma_m} \\
&& \times R_{\mu_1\nu_1}^{\phantom{\mu_1\nu_1}\lambda_1\sigma_1} R_{\mu_2\nu_2}^{\phantom{\mu_2\nu_2}\lambda_2\sigma_2} \ldots R_{\mu_m\nu_m}^{\phantom{\mu_m\nu_m}\lambda_m\sigma_m}\,.
\end{eqnarray}

The antisymmetric tensor is nonzero only when the indices $\mu,\mu_1,\nu_1,\mu_2,\nu_2,\ldots\mu_m,\nu_m$ are all distinct. Thus, the general Lovelock theory is such that $2\m <D$. In particular, for $D=4$, we have $\m=1$ corresponding to the Einstein theory \cite{Lovelock:1972vz}. When $D=5$ or $6$, $\m=2$ and one has the (quadratic in curvature) Einstein-Gauss-Bonnet theory with the coupling constant $\alpha_2$.

Following \cite{Konoplya:2017lhs}, we introduce
\begin{equation}\label{amdef}
\a_m=\frac{\alpha_m}{m}\frac{(D-3)!}{(D-2m-1)!}=\frac{\alpha_m}{m}\prod_{p=1}^{2m-2}(D-2-p)
\end{equation}
and consider the limit $D\to 4$ while $\a_m$ remain constant. In this way, we obtain the regularized $4D$ Einstein-Lovelock theory formulated in \cite{Konoplya:2020qqh}, which generalizes the approach of \cite{Glavan:2019inb} used for the Einstein-Gauss-Bonnet theory.

The four-dimensional static and spherically symmetric black hole in the Einstein-Lovelock theory ($\Lambda=0$) is given by the following line element \cite{Konoplya:2020qqh}:
\begin{equation}\label{Lmetric}
  ds^2=-f(r)dt^2+\frac{1}{f(r)}dr^2 + r^2 (d\theta^2+d \sin^2\theta d\phi^2).
\end{equation}
The metric function $f(r)$ is defined through a new variable $\psi(r)$,
\begin{equation}\label{Lfdef}
f(r)=1-r^2\,\psi(r),
\end{equation}
which satisfies the algebraic equation \cite{Takahashi:2010}
\begin{equation}\label{MEq}
W[\psi(r)]\equiv\psi(r)+\sum_{m=2}^{\m}\a_m\psi(r)^m=\frac{2M}{r^{3}}\,,
\end{equation}
where $M$ is the asymptotic mass \cite{Myers:1988ze}, $\m$ is the power of curvature of the corresponding Lovelock term. The Lovelock corrections result in more than one branch, only one of which is perturbative in $\a_m$ for the Einstein-Lovelock theory of any order.

For example, the (quadratic in curvature, so $\m=2$) Gauss-Bonnet theory  leads to the two branches \cite{Fernandes:2020rpa}:
\begin{equation}
  f(r)=1-\frac{r^2}{2\a_2}\left(-1\pm\sqrt{1+\frac{8\a_2M}{r^3}}\right),
\end{equation}
one of which, corresponding to the ``+'' sign, is perturbative in $\a_2$, while for the ``-'' the metric function $f(r)$ goes to infinity when $\a_2 \to 0$.

It is convenient to measure all dimensional quantities in units of the horizon radius $r_0$. For the asymptotic mass we obtain
\begin{equation}\label{Mdef}
  2M=r_0\left(1+\sum_{m=2}^{\m}\frac{\a_m}{r_0^{2m-2}}\right).
\end{equation}

The metric function $f(r)$ for the perturbative branch of the general Einstein-Lovelock black hole can be obtained numerically \cite{Konoplya:2020qqh}.\footnote{The Mathematica\textregistered{} code for the metric-function calculation is available from \url{https://arxiv.org/src/2003.07788/anc/}.}

\section{Cut-off due to the eikonal instability}\label{sec:cutoff}

In order to understand how important the Lovelock correction at $i$-th order in curvature is, let us first make here one observation about black-hole stability at different orders of Lovelock theory. Let us suppose that the coupling constants $\a_m$ are either
a) all positive or null or
b) all negative or null,
that is, the coupling constants at different orders cannot be of opposite signs. In the most general framework this is certainly not a strict supposition, unless one associates the coupling constant with the fundamental string scale. Therefore, later we will also discuss the case in which $\a_m$ can be both positive and negative at different orders.

Let us consider the Einstein-Lovelock theory with the only one non-zero Lovelock term, that is, all $\a_{m\neq i} =0$, except one $\a_{i}>0$, and designate the threshold value of this coupling constant  $\a_{i}=\a_{i}^{crit}$ at which the eikonal instability \cite{Konoplya:2020juj} occurs. Then, if one or more other Lovelock terms are added to this system, that is, more coupling constants are turned on $\a_m>0$, then the critical value of the $\a_{i}$ will always decrease. This is an important observation, because it means that in order to estimate the maximal deformation of the black-hole geometry caused by the $i-th$ order Lovelock term, it is sufficient to consider all other Lovelock corrections as vanishing $\a_{m \neq i} = 0$. Whenever more Lovelock terms are added, the relative deformation of the geometry caused by a given term will only decrease. Here the measure of deformation or deviation of the geometry from its Schwarzschild limit is understood as a deviation of some gauge invariant observable quantity, such as radius of the shadow or frequency at the innermost stable circular orbit (ISCO).

Now, we are in position to discuss observable quantities, which we use for measuring deviation of the $4D$ Einstein-Lovelock black holes from the Schwarzschild geometry: quasinormal modes, radius of the shadow, Lyapunov exponent, and the frequency at the innermost stable orbit.

\subsection{Quasinormal modes}
For the purpose of illustration here we will mainly study a test electromagnetic field, although the analysis can be easily extended to the gravitational perturbations, the effective potentials for which were obtained in \cite{Konoplya:2020juj}.
The general covariant equations for an electromagnetic field has the form
\begin{equation}\label{EmagEq}
\frac{1}{\sqrt{-g}}\partial_\mu \left(F_{\rho\sigma}g^{\rho \nu}g^{\sigma \mu}\sqrt{-g}\right)=0\,,
\end{equation}
where $F_{\rho\sigma}=\partial_\rho A_{\sigma}-\partial_\sigma A_{\rho}$ and $A_\mu$ is a vector potential.
After separation of the variables the perturbation equation~(\ref{EmagEq}) takes the following general wave-like form
\begin{equation}\label{wave-equation}
\left(\frac{d^2}{dr_*^2}+\omega^2-f(r)\frac{\ell(\ell+1)}{r^2}\right)\Psi=0,
\end{equation}
where $\ell$ is the multipole number, and $r_*$ is the ``tortoise'' coordinate, defined as $$dr_*=\frac{dr}{f(r)}.$$

Quasinormal modes $\omega_{n}$ correspond to solutions of the master wave equation (\ref{wave-equation}) with the requirement of the purely outgoing waves at infinity and purely incoming waves at the event horizon:
\begin{equation}
\Psi \propto e^{\pm i \omega r^{*}}, \quad r^{*} \to \pm \infty.
\end{equation}
In order to find quasinormal modes here we will use the sixth order WKB formula developed in \cite{Schutz:1985km,Iyer:1986np,Konoplya:2003ii} (see also \cite{Konoplya:2019hlu,Konoplya:2011qq} for reviews) and time-domain integration, proposed in \cite{Gundlach:1993tp}.

\begin{table*}
\begin{tabular}{|c|c|c|c|c|c|c|}
  \hline
  order & $\alpha_{inst}$ & $R_{sh}$ & $\lambda$ & $\Omega_{ISCO}$ & $\omega$ (QNM) (1st order) &  $\omega$ (QNM) (6th order)  \\
 \hline
  2  & 0.0890~~ & 2.76248 & 0.335675 & 0.130467 & $0.536410 - 0.160180 i$ & $0.471821 - 0.163727 i$ \\
 \hline
  3  & 0.0145~~ & 2.63255 & 0.375673 & 0.134254 & $0.566013 - 0.178276 i$ & $0.494739 - 0.176565 i$ \\
 \hline
  4  & 0.0050~~ & 2.61074 & 0.382160 & 0.135409 & $0.571197 - 0.181209 i$ & $0.496803 - 0.178715 i$ \\
 \hline
  5  & 0.0023~~ & 2.60401 & 0.383823 & 0.135771 & $0.572766 - 0.181969 i$ & $0.492613 - 0.183712 i$ \\
 \hline
  6  & 0.001281 & 2.60140 & 0.384359 & 0.135909 & $0.573361 - 0.182216 i$ & $0.490779 - 0.187169 i$ \\
 \hline
  7  & 0.000776 & 2.60009 & 0.384589 & 0.135977 & $0.573654 - 0.182324 i$ & $0.492773 - 0.187527 i$ \\
 \hline
  8  & 0.000500 & 2.59937 & 0.384705 & 0.136015 & $0.573814 - 0.182378 i$ & $0.495693 - 0.186350 i$ \\
 \hline
  9  & 0.000348 & 2.59898 & 0.384765 & 0.136035 & $0.573901 - 0.182407 i$ & $0.497424 - 0.185287 i$ \\
 \hline
  10 & 0.000250 & 2.59873 & 0.384804 & 0.136049 & $0.573957 - 0.182425 i$ & $0.497769 - 0.184813 i$ \\
 \hline
  11 & 0.000185 & 2.59856 & 0.384829 & 0.136058 & $0.573995 - 0.182437 i$ & $0.497362 - 0.184795 i$ \\
 \hline
  12 & 0.000141 & 2.59844 & 0.384846 & 0.136064 & $0.574020 - 0.182445 i$ & $0.496825 - 0.184954 i$ \\
 \hline
  13 & 0.000110 & 2.59836 & 0.384858 & 0.136068 & $0.574038 - 0.182451 i$ & $0.496442 - 0.185116 i$ \\
 \hline
  14 & 0.000087 & 2.59830 & 0.384867 & 0.136071 & $0.574051 - 0.182455 i$ & $0.496256 - 0.185221 i$ \\
 \hline
  15 & 0.000070 & 2.59826 & 0.384873 & 0.136073 & $0.574061 - 0.182458 i$ & $0.496207 - 0.185272 i$ \\
 \hline
  16 & 0.000058 & 2.59823 & 0.384878 & 0.136075 & $0.574068 - 0.182460 i$ & $0.496223 - 0.185288 i$ \\
 \hline
  17 & 0.000048 & 2.59820 & 0.384882 & 0.136076 & $0.574073 - 0.182462 i$ & $0.496262 - 0.185288 i$ \\
 \hline
  18 & 0.000040 & 2.59818 & 0.384885 & 0.136077 & $0.574078 - 0.182463 i$ & $0.496300 - 0.185284 i$ \\
 \hline
  19 & 0.000034 & 2.59816 & 0.384887 & 0.136078 & $0.574081 - 0.182464 i$ & $0.496328 - 0.185279 i$ \\
  \hline\hline
  \multicolumn{2}{|l|}{1 (Schwarzschild)}
                         & 2.59808 & 0.384900 & 0.136083 & $0.574101 - 0.182471 i$ & $0.496383 - 0.185274 i$ \\
  \hline
\end{tabular}
\caption{The threshold value of (in)stability $\alpha_{inst}$, radius of the shadow $R_{sh}$, Lyapunov exponent $\lambda$, frequency at ISCO $\Omega_{ISCO}$ and the fundamental quasinormal mode $\omega$ for electromagnetic perturbations ($\ell=1$, $n=0$) calculated by the 6th order WKB formula. All quantities are measured in units of the event horizon radius ($r_0=1$); the last line corresponds to the Schwarzschild black hole.}\label{tabl:positive}
\end{table*}

\begin{table}
\begin{tabular}{|c|c|c|c|c|}
  \hline
  order &  $R_{sh}$ & $\lambda$ & $\Omega_{ISCO}$ & fundamental QNM \\
 \hline
  2 &  1.70722 & 1.054030 & 0.166374 & $0.537-0.388i$ \\
 \hline
  3 &  1.87218 & 0.819374 & 0.188296 & $0.539-0.347i$ \\
 \hline
  4 &  1.99829 & 0.667574 & 0.180030 & $0.541-0.321i$ \\
 \hline
  5 &  2.09629 & 0.564746 & 0.169830 & $0.541-0.302i$ \\
 \hline
  6 &  2.17223 & 0.500088 & 0.163101 & $0.541-0.287i$ \\
 \hline
  7 &  2.23081 & 0.463993 & 0.158578 & $0.541-0.275i$ \\
 \hline
  8 &  2.27627 & 0.445002 & 0.155346 & $0.540-0.265i$ \\
 \hline\hline
  1 &  2.59808 & 0.384900 & 0.136083 & $0.496-0.185i$ \\
 \hline
\end{tabular}
\caption{Radius of the shadow $R_{sh}$, Lyapunov exponent $\lambda$, frequency at ISCO $\Omega_{ISCO}$ and the fundamental quasinormal mode or electromagnetic perturbations ($\ell=1$, $n=0$) calculated by the time-domain integration; for each order $i$ the only nonzero coupling constant has the near-extreme negative value, $\a_{m=i} = -m^{-1} + 10^{-3}$. All quantities are measured in units of the event horizon radius ($r_0=1$); the last line corresponds to the Schwarzschild black hole.}\label{tabl:threshold}
\end{table}

From Table~\ref{tabl:positive} for the positive values of coupling constants we can see that the quasinormal modes for the third order Einstein-Lovelock (with the vanishing Gauss-Bonnet term) black hole differs from that for the purely Einstein theory by about four percents, while already at the fifth order the difference is only slightly exceeds one percent.

From here we conclude that only first few Lovelock correction may change the quasinormal modes seemingly, while higher than the 5th order can be safely ignored. This is not so for the case of negative $\a_{m}$ for which, as can be seen in Table~\ref{tabl:threshold}, even the 8th Lovelock order change the quasinormal frequency by more than ten percents. Nevertheless, the convergence in Lovelock orders takes place even in this case and each higher orders contributes less owing to the constrain upon the values of the negative coupling constants. Notice that WKB method does not provide reliable results for large negative values of the Lovelock coupling. That is why we have calculated the dominant QNMs by fitting time-domain profiles.

As can be seen from the comparison of the first-order WKB formula, which depends on the first two derivatives of the metric function, and a more accurate sixth-order WKB formula, which depends on higher (up to 12th) derivatives, in order to estimate the order of deviation from the Schwarzschild geometry it sufficient to consider only quantities, which depend on the lowest derivatives. Since the first-order WKB formula represent the eikonal regime, in the next subsection we shall see that the order of effect due to the Lovelock terms on the fundamental quasinormal modes can be estimated by considering the shadow radius and the Lyapunov exponent.

\subsection{Shadow radius and the Lyapunov exponent}
First we need to find the radius of the photon sphere $r_{ph}$, which for a spherically symmetric solution is determined by means of the following function: (see, for example, \cite{Bisnovatyi-Kogan:2017kii})
\begin{equation}
h^2(r) \equiv \frac{r^2}{f(r)} \,.
\label{h2 definition}
\end{equation}
The photon sphere corresponds to the minimum of $h(r)$, so that in order to calculate $r_{ph}$ we find the solution to the equation
\begin{equation}
h'(r)=0\,.
\end{equation}
Then, the radius of the black-hole shadow $R_{sh}$, as seen by a distant static observer located at $r_O$, obeys
\begin{equation}
R_{sh} = \frac{h(r_{ph})r_O}{h(r_O)} = \frac{r_{ph}\sqrt{f(r_O)}}{\sqrt{f(r_{ph})}} \approx  \frac{r_{ph}}{\sqrt{f(r_{ph})}}\,,
\label{shadow def}
\end{equation}
where we assumed that the observer is located sufficiently far away from the black hole so that $f(r_O) \approx 1$.

The Lyapunov exponent $\lambda$ characterizes the mean lifetime of particles at the photon sphere, and depends on the second derivative of the metric function at $r=r_{ph}$,
\begin{equation}
\lambda = r_{ph}^2\sqrt{\frac{h''(r_{ph})}{h^5(r_{ph})}}=\frac{f(r_{ph})}{r_{ph}}\sqrt{1-\frac{r_{ph}^2f''(r_{ph})}{2f(r_{ph})}}.
\end{equation}

In the eikonal limit $\ell \to \infty$ the quasinormal modes of test fields are connected with the parameters of null geodesics, namely, the damping rate is related with the Lyapunov exponent $\lambda$ and the real oscillation frequency is related with the orbital frequency \cite{Cardoso:2008bp},
\begin{equation}
\omega=\frac{1}{R_{sh}}\left(\ell+\frac{1}{2}\right)-i\lambda\left(n+\frac{1}{2}\right)+{\cal O}\left(\frac{1}{\ell}\right).
\end{equation}
This correspondence, however not guaranteed for gravitational field, is always valid for minimally coupled test fields \cite{Konoplya:2017wot}.

From Table~\ref{tabl:positive} we can see that for positive coupling constants the radius of the shadow and the Lyapunov exponent already at the fourth Lovelock order differs from the Schwarzschild case by less than one percent. Therefore, for this case, we can safely ignore the fifth and higher Lovelock orders as it is unlikely that the deviation from the Einstein theory for the geodesics' parameters could be observed with accuracy of $\sim 0.2\%$ in the near future. On the contrary, for negative coupling constants (see Table~\ref{tabl:threshold}) even at the 8th Lovelock order the shadow radius and the Lyapunov exponent differ from the Schwarzschild values by tens of percents.

\subsection{Frequencies at the innermost stable circular orbit}
The condition of the circular orbit is
\begin{equation}
V_{eff}(r) = V'_{eff}(r) =0,
\end{equation}
where $V_{eff}$ is the effective potential for a particle of unit mass with energy $E$ and angular momentum $L$,
\begin{equation}
V_{eff}(r) = \frac{E^2}{f(r)} -\frac{L^2}{r^2} -1.
\end{equation}
The innermost stable circular orbit corresponds to vanishing of the second derivative of the effective potential:
\begin{equation}
V_{eff}''(r_{ISCO})=0.
\end{equation}
The frequency at ISCO is
\begin{equation}
\Omega_{ISCO} = \frac{d \phi}{d t} \bigg|_{r= r_{ISCO}} = \sqrt{\frac{f'(r)}{2 r}}\bigg|_{r= r_{ISCO}}\,.
\end{equation}

In the case of positive coupling constant the correction due to the fourth order Lovelock term is less than one percent, so that the fifth and higher orders could practically ignored. The negative couplings, again, are characterized by very slow convergence of Lovelock orders with the effect of tens of percent even at the 8th order (see Table~\ref{tabl:threshold}).

If take a completely agnostic position and suppose that coupling constants can change the sign from one Lovelock order to another, then the negative couplings have opposite effect to the positive ones: they enlarge the region os stability and make slower the convergence of Lovelock terms. Anyway, there are constrains on these coupling constants and convergence takes place even for mixed (positive and negative) $\a_{m}$.

\section{Parameterized description of $4D$ Einstein-Lovelock black holes}\label{sec:param}

In \cite{Konoplya:2020hyk} it was shown that the so called moderate metrics can be approximated by the following parametrization:
\begin{equation}\label{parametrization}
f(r)=1 - \frac{r_{0}(\epsilon+1)}{r} + \frac{r_{0}^3(\epsilon+a_1)}{r^3} - \frac{r_0^4 a_{1}}{r^4},
\end{equation}

Here $r_0$ is the radius of the event horizon, so that $N(r_0)=0$; $\epsilon$, $a_1$ are some parameters, such that the Schwarzschild limit is reproduced when they all are equal to zero. Moderate metrics are those which can be approximated by the above parametrization with sufficient accuracy. Normally they include a class of metrics, for which the metric functions change relatively slowly from the event horizon up to the end of the effective zone of radiation, which is about the innermost stable circular orbit. This allows one to describe, in a unified way, black-hole metrics whose observable values are distinguishable from those of the Schwarzschild solution \cite{Konoplya:2020hyk}.

For the Einstein-Lovelock theory we have
\begin{eqnarray}
\epsilon &=& \frac{2M}{r_0}-1 = r_0^2W\left[r_0^{-2}\right]-1 = \sum_{m=2}^{\m}\frac{\a_m}{r_0^{2m-2}},
\\
a_1&=&2\epsilon+4\pi r_0 T_H-1=r_0^2W\left[r_0^{-2}\right]\left(2+\frac{3}{W'\left[r_0^{-2}\right]}\right)-5
\nonumber\\&=&2\sum_{m=2}^{\m}\frac{\a_m}{r_0^{2m-2}}-3\frac{\sum_{m=2}^{\m}(m-1)\a_mr_0^{-2m}}{\sum_{m=2}^{\m}m\a_mr_0^{-2m}+r_0^{-2}},
\end{eqnarray}
where $M$ is the asymptotic mass and $T_H$ is the Hawking temperature.

When this parametrization via only the two parameters $\epsilon$ and $a_1$ is not a sufficiently good approximation, we can use the general parametrization developed in  \cite{Rezzolla:2014mua}. Namely, we approximate the metric function by a rational function of $r$, which is introduced through the continued fraction as
\begin{equation}
f(r)=\frac{r-r_0}{r}\left(1-\frac{\epsilon r_0}{r}-\frac{\epsilon r_0^2}{r^2}+\dfrac{\dfrac{a_1r_0^3}{r^3}}{1+\dfrac{a_2\dfrac{r-r_0}{r}}{1+\ldots}}\right),
\end{equation}
where dimensionless constants $a_1,a_2,\ldots$ are fixed by matching the series expansion for $f(r)$ at the event horizon. In particular, one can find that
\begin{widetext}
\begin{equation}
a_2=-3\frac{r_0^2(3W\left[r_0^{-2}\right]^2W''\left[r_0^{-2}\right]+2W\left[r_0^{-2}\right]W'\left[r_0^{-2}\right]^3+8W\left[r_0^{-2}\right]W'\left[r_0^{-2}\right]^2)-10W'\left[r_0^{-2}\right]^3}{2r_0^2W\left[r_0^{-2}\right]W'\left[r_0^{-2}\right]^2(2W'\left[r_0^{-2}\right]+3)-10W'\left[r_0^{-2}\right]^3}.
\end{equation}
\end{widetext}

The Mathematica\textregistered{} code, which constructs the approximation of any given order for the metric function for arbitrary values of the Lovelock coupling constants $\a_2,\a_3,\ldots$, is attached to the arXiv preprint as a supplementary material.\footnote{The Mathematica\textregistered{} code for the approximate metric functions is available from \url{https://arxiv.org/src/2005.02225/anc/}.}

\begin{table}
\begin{tabular}{|c|c|c|c|c|}
  \hline
  quantity & value  & effect  & 1st order error & 2d order error  \\
 \hline
 $R_{sh}$  & 2.380080 & ~8.39 \%  & 0.429 \%  & 0.198 \%    \\
 \hline
  $\lambda$ & 0.470996 & 22.37 \%  & 1.956 \%  & 1.153 \%  \\
 \hline
 $\Omega_{ISCO}$ & 0.143640 & ~5.55 \%  & 0.906 \%  & 0.542 \%  \\
  \hline
\end{tabular}
\caption{The shadow radius $R_{sh}$, Lyapunov exponent $\lambda$, and frequency at ISCO $\Omega_{ISCO}$
for $r_0=1$, $\a_{2} = -0.12$, $\a_{3} = -0.011$, $\a_{4} = 0.009$, $\a_{5} = 0.002$, $\a_{6} = -0.001$. }\label{tabl:param}
\end{table}

In Table~\ref{tabl:param} we give an example of description of the six parameter black hole, depending upon the five Lovelock coupling constants and mass, in terms of only two ($\epsilon$ and $a_1$) or three (when $a_2$ is added) parameters of the parametrization (\ref{parametrization}). From Table~\ref{tabl:param} we can see not only the values of observable quantities, but also the corresponding deviations from their Schwarzschild limit when calculated for the parametrized black hole at the first ($\epsilon$ and $a_1$ are non-zero) and second ($a_2$ is not zero as well) order of the parametrization. It is natural to assume that if the effect is at least one order larger than the error, the approximation is sufficiently accurate. From the above example for the particular values of the coupling constants we see that while the shadow radius 
and the Lyapunov exponent can be calculated with the relative error which is one order smaller than the effect already at the first order of the parametrization, this is not so for the frequency at the innermost stable orbit, for which the effect is only 5 times larger than the error. The second order remedy this situation.

In general, we observe that a) the three parameter ($\epsilon$, $a_1$ and $a_2$) approximation (\ref{parametrization}) is sufficient to represent $4D$ Einstein-Lovelock black hole with many coupling constants for the most part of the range of values of the coupling constants $\a_{m}$, if the black hole is sufficiently far from the threshold of instability and b) when the coupling constants are relatively small and decreasing by absolute value when going over to higher orders of the Lovelock series, the description via the two parameters only ($\epsilon$ and $a_1$) gives the relative error at least one order smaller than the effect. Therefore the parametrization converges very fast, when we are limited by strictly positive coupling constants because of the natural cut-off on the values the higher order couplings.

\section{Conclusions}
Here we have studied properties of the $4D$ Einstein-Lovelock black holes obtained as a result of the dimensional regularization suggested in \cite{Casalino:2020kbt,Konoplya:2020qqh} in a similar fashion with the recent approach for the $4D$ Einstein-Gauss-Bonnet black hole \cite{Glavan:2019inb}. In particularly, we have learned what is the role of higher curvature corrections given by the Lovelock terms and how to describe the multi-parameter black holes when a lot of Lovelock terms are taken into consideration, so that analysis of various effects on every parameter would be very cumbersome problem. When we are limited by positive values of the coupling constants, the eikonal instability of the black hole gives the answer to the first question: higher order Lovelock terms have swiftly decreasing upper bound on the absolute value of the corresponding coupling constant, so that its influence decreases quickly and only a few first Lovelock terms can lead to noticeable effect on the geometry, while higher order can be safely ignored. When negative couplings are allowed, this is not so, despite influence of higher orders are smaller as well.
If implying the low-energy limit description of small black holes with positive coupling constants, when decreasing the black-hole size, higher order Lovelock corrections should play more important role, as the corresponding coupling constant is divided by the radius of the black hole at some power. At the same the instability will cut off the values of the coupling constant, so that a few first order of the Lovelock series will be always enough for an adequate description of even small black holes.
This remarkable results does not take place in the $D>4$ Einstein-Lovelock theory for which the instability does not cut-off higher orders of the Lovelock series  \cite{Konoplya:2017lhs}.

We also find an elegant answer to the question how to treat the black hole geometry when many coupling constants due to Lovelock terms come into play. Using the general parametrization for the arbitrary spherically symmetric black holes we find that approximate description of the black hole geometry is possible with only two or three parameters when the black hole is sufficiently far from the instability threshold.

\begin{acknowledgments}
The authors acknowledge the support of the grant 19-03950S of Czech Science Foundation (GAČR). This publication has been prepared with partial support of the ``RUDN University Program 5-100'' (R. K.).
\end{acknowledgments}


\begin{thebibliography}{99}
\bibitem{Lovelock:1971yv}
  D.~Lovelock,
  J.\ Math.\ Phys.\  {\bf 12}, 498 (1971)
  doi:10.1063/1.1665613;

\bibitem{Lovelock:1972vz}
  D.~Lovelock,
  J.\ Math.\ Phys.\  {\bf 13}, 874 (1972)
  doi:10.1063/1.1666069.

\bibitem{Konoplya:2010vz}
  R.~A.~Konoplya and A.~Zhidenko,
  Phys.\ Rev.\ D {\bf 82}, 084003 (2010)
  doi:10.1103/PhysRevD.82.084003
  [arXiv:1004.3772 [hep-th]].

\bibitem{Dotti:2005sq}
  G.~Dotti and R.~J.~Gleiser,
  Phys.\ Rev.\ D {\bf 72}, 044018 (2005)
  doi:10.1103/PhysRevD.72.044018
  [gr-qc/0503117].

\bibitem{Gleiser:2005ra}
  R.~J.~Gleiser and G.~Dotti,
  Phys.\ Rev.\ D {\bf 72}, 124002 (2005)
  doi:10.1103/PhysRevD.72.124002
  [gr-qc/0510069].

\bibitem{Konoplya:2008ix}
  R.~A.~Konoplya and A.~Zhidenko,
  Phys.\ Rev.\ D {\bf 77}, 104004 (2008)
  doi:10.1103/PhysRevD.77.104004
  [arXiv:0802.0267 [hep-th]].

\bibitem{Takahashi:2011qda}
  Prog.\ Theor.\ Phys.\  {\bf 125}, 1289 (2011)
  doi:10.1143/PTP.125.1289
  [arXiv:1102.1785 [gr-qc]].

\bibitem{Takahashi:2012np}
  T.~Takahashi,
  PTEP {\bf 2013}, 013E02 (2013)
  doi:10.1093/ptep/pts049
  [arXiv:1209.2867 [gr-qc]].

\bibitem{Yoshida:2015vua}
  D.~Yoshida and J.~Soda,
  Phys.\ Rev.\ D {\bf 93}, no. 4, 044024 (2016)
  doi:10.1103/PhysRevD.93.044024
  [arXiv:1512.05865 [gr-qc]].

\bibitem{Reall:2014pwa}
H.~Reall, N.~Tanahashi and B.~Way,
Class. Quant. Grav. \textbf{31}, 205005 (2014)
doi:10.1088/0264-9381/31/20/205005
[arXiv:1406.3379 [hep-th]].

\bibitem{Cuyubamba:2016cug}
  M.~A.~Cuyubamba, R.~A.~Konoplya and A.~Zhidenko,
  Phys.\ Rev.\ D {\bf 93}, no. 10, 104053 (2016)
  doi:10.1103/PhysRevD.93.104053
  [arXiv:1604.03604 [gr-qc]].

\bibitem{Konoplya:2017lhs}
  R.~A.~Konoplya and A.~Zhidenko,
  JCAP {\bf 1705}, 050 (2017)
  doi:10.1088/1475-7516/2017/05/050
  [arXiv:1705.01656 [hep-th]].

\bibitem{Konoplya:2017zwo}
  R.~A.~Konoplya and A.~Zhidenko,
  JHEP {\bf 1709}, 139 (2017)
  doi:10.1007/JHEP09(2017)139
  [arXiv:1705.07732 [hep-th]].

\bibitem{Cardoso:2008bp}
  V.~Cardoso, A.~S.~Miranda, E.~Berti, H.~Witek and V.~T.~Zanchin,
  Phys.\ Rev.\ D {\bf 79}, 064016 (2009)
  doi:10.1103/PhysRevD.79.064016
  [arXiv:0812.1806 [hep-th]].

\bibitem{Konoplya:2017wot}
  R.~A.~Konoplya and Z.~Stuchlík,
  Phys.\ Lett.\ B {\bf 771}, 597 (2017)
  doi:10.1016/j.physletb.2017.06.015
  [arXiv:1705.05928 [gr-qc]].

\bibitem{Glavan:2019inb}
  D.~Glavan and C.~Lin,
  Phys.\ Rev.\ Lett.\  {\bf 124}, no. 8, 081301 (2020)
  doi:10.1103/PhysRevLett.124.081301
  [arXiv:1905.03601 [gr-qc]].

\bibitem{Tomozawa:2011gp}
  Y.~Tomozawa,
  arXiv:1107.1424 [gr-qc].

\bibitem{Casalino:2020kbt}
  A.~Casalino, A.~Colleaux, M.~Rinaldi and S.~Vicentini,
  arXiv:2003.07068 [gr-qc].

\bibitem{Konoplya:2020qqh}
R.~Konoplya and A.~Zhidenko,
Phys. Rev. D \textbf{101}, no.8, 084038 (2020)
doi:10.1103/PhysRevD.101.084038
[arXiv:2003.07788 [gr-qc]].

\bibitem{Konoplya:2020bxa}
  R.~A.~Konoplya and A.~F.~Zinhailo,
  arXiv:2003.01188 [gr-qc].

\bibitem{Guo:2020zmf}
  M.~Guo and P.~C.~Li,
  arXiv:2003.02523 [gr-qc].

\bibitem{Fernandes:2020rpa}
P.~G.~S.~Fernandes,
Phys. Lett. B \textbf{805}, 135468 (2020)
doi:10.1016/j.physletb.2020.135468
[arXiv:2003.05491 [gr-qc]].

\bibitem{Wei:2020ght}
  S.~W.~Wei and Y.~X.~Liu,
  arXiv:2003.07769 [gr-qc].

\bibitem{Kumar:2020owy}
  R.~Kumar and S.~G.~Ghosh,
  arXiv:2003.08927 [gr-qc].

\bibitem{Hegde:2020xlv}
  K.~Hegde, A.~N.~Kumara, C.~L.~A.~Rizwan, A.~K.~M. and M.~S.~Ali,
  arXiv:2003.08778 [gr-qc].

\bibitem{Zhang:2020qew}
  Y.~P.~Zhang, S.~W.~Wei and Y.~X.~Liu,
  arXiv:2003.10960 [gr-qc].

\bibitem{Lu:2020iav}
  H.~Lu and Y.~Pang,
  arXiv:2003.11552 [gr-qc].

\bibitem{Konoplya:2020ibi}
  R.~A.~Konoplya and A.~Zhidenko,
  arXiv:2003.12171 [gr-qc].

\bibitem{Ghosh:2020syx}
  S.~G.~Ghosh and R.~Kumar,
  arXiv:2003.12291 [gr-qc].

\bibitem{Konoplya:2020juj}
  R.~A.~Konoplya and A.~Zhidenko,
  arXiv:2003.12492 [gr-qc].

\bibitem{Kobayashi:2020wqy}
  T.~Kobayashi,
  arXiv:2003.12771 [gr-qc].

\bibitem{Zhang:2020qam}
  C.~Y.~Zhang, P.~C.~Li and M.~Guo,
  arXiv:2003.13068 [hep-th].

\bibitem{Kumar:2020uyz}
  A.~Kumar and R.~Kumar,
  arXiv:2003.13104 [gr-qc].

\bibitem{HosseiniMansoori:2020yfj}
  S.~A.~Hosseini Mansoori,
  arXiv:2003.13382 [gr-qc].

\bibitem{Wei:2020poh}
S.~W.~Wei and Y.~X.~Liu,
Phys. Rev. D \textbf{101}, no.10, 104018 (2020)
doi:10.1103/PhysRevD.101.104018
[arXiv:2003.14275 [gr-qc]].

\bibitem{Singh:2020nwo}
  D.~V.~Singh, S.~G.~Ghosh and S.~D.~Maharaj,
  arXiv:2003.14136 [gr-qc].

\bibitem{Churilova:2020aca}
  M.~S.~Churilova,
  arXiv:2004.00513 [gr-qc].

\bibitem{Mishra:2020gce}
  A.~K.~Mishra,
  arXiv:2004.01243 [gr-qc].

\bibitem{Heydari-Fard:2020sib}
  M.~Heydari-Fard, M.~Heydari-Fard and H.~R.~Sepangi,
  arXiv:2004.02140 [gr-qc].

\bibitem{Konoplya:2020cbv}
  R.~A.~Konoplya and A.~F.~Zinhailo,
  arXiv:2004.02248 [gr-qc].

\bibitem{Jin:2020emq}
  X.~H.~Jin, Y.~X.~Gao and D.~J.~Liu,
  arXiv:2004.02261 [gr-qc].

\bibitem{Zhang:2020sjh}
  C.~Y.~Zhang, S.~J.~Zhang, P.~C.~Li and M.~Guo,
  arXiv:2004.03141 [gr-qc].

\bibitem{EslamPanah:2020hoj}
B.~Eslam Panah and K.~Jafarzade,
[arXiv:2004.04058 [hep-th]].

\bibitem{NaveenaKumara:2020rmi}
  A.~Naveena Kumara, C.~L.~A.~Rizwan, K.~Hegde, M.~S.~Ali and A.~K.~M,
  arXiv:2004.04521 [gr-qc].

\bibitem{Aragon:2020qdc}
  A.~Aragón, R.~Bécar, P.~A.~González and Y.~Vásquez,
  arXiv:2004.05632 [gr-qc].

\bibitem{Malafarina:2020pvl}
  D.~Malafarina, B.~Toshmatov and N.~Dadhich,
  arXiv:2004.07089 [gr-qc].

\bibitem{Yang:2020czk}
  S.~J.~Yang, J.~J.~Wan, J.~Chen, J.~Yang and Y.~Q.~Wang,
  arXiv:2004.07934 [gr-qc].

\bibitem{Fernandes:2020nbq}
  P.~G.~S.~Fernandes, P.~Carrilho, T.~Clifton and D.~J.~Mulryne,
  arXiv:2004.08362 [gr-qc].

\bibitem{Cuyubamba:2020moe}
  M.~A.~Cuyubamba,
  arXiv:2004.09025 [gr-qc].

\bibitem{Mahapatra:2020rds}
  S.~Mahapatra,
  arXiv:2004.09214 [gr-qc].

\bibitem{Shu:2020cjw}
  F.~W.~Shu,
  arXiv:2004.09339 [gr-qc].

\bibitem{Casalino:2020pyv}
  A.~Casalino and L.~Sebastiani,
  arXiv:2004.10229 [gr-qc].

\bibitem{Liu:2020evp}
  P.~Liu, C.~Niu and C.~Y.~Zhang,
  arXiv:2004.10620 [gr-qc].

\bibitem{Devi:2020uac}
  S.~Devi, R.~Roy and S.~Chakrabarti,
  arXiv:2004.14935 [gr-qc].

\bibitem{Ma:2020ufk}
  L.~Ma and H.~Lu,
  arXiv:2004.14738 [gr-qc].

\bibitem{Liu:2020yhu}
  P.~Liu, C.~Niu, X.~Wang and C.~Y.~Zhang,
  arXiv:2004.14267 [gr-qc].

\bibitem{Kumar:2020sag}
  R.~Kumar, S.~U.~Islam and S.~G.~Ghosh,
  arXiv:2004.12970 [gr-qc].

\bibitem{Churilova:2020mif}
  M.~S.~Churilova,
  arXiv:2004.14172 [gr-qc].

\bibitem{Ge:2020tid}
  X.~H.~Ge and S.~J.~Sin,
  arXiv:2004.12191 [hep-th].

\bibitem{Zeng:2020dco}
  X.~X.~Zeng, H.~Q.~Zhang and H.~Zhang,
  arXiv:2004.12074 [gr-qc].

\bibitem{Gurses:2020ofy}
  M.~Gurses, T.~C.~Sisman and B.~Tekin,
  arXiv:2004.03390 [gr-qc].

\bibitem{Tian:2020nzb}
  S.~X.~Tian and Z.~H.~Zhu,
  arXiv:2004.09954 [gr-qc].

\bibitem{Arrechea:2020evj}
  J.~Arrechea, A.~Delhom and A.~Jiménez-Cano,
  arXiv:2004.12998 [gr-qc].

\bibitem{Cognola}
G.~Cognola, R.~Myrzakulov, L.~Sebastiani and S.~Zerbini,
  Phys.\ Rev.\ D {\bf 88}, no. 2, 024006 (2013)
  doi:10.1103/PhysRevD.88.024006
  [arXiv:1304.1878 [gr-qc]].

\bibitem{Cai:2009ua}
  R.~G.~Cai, L.~M.~Cao and N.~Ohta,
  JHEP {\bf 1004}, 082 (2010)
  doi:10.1007/JHEP04(2010)082
  [arXiv:0911.4379 [hep-th]].

\bibitem{Hennigar:2020}
R.~A.~Hennigar, D.~Kubiznak, R.~B.~Mann and C.~Pollack,
[arXiv:2004.09472 [gr-qc]];
  arXiv:2004.12995 [gr-qc].

\bibitem{Lu:2020mjp}
H.~Lu and P.~Mao,
[arXiv:2004.14400 [hep-th]].
  
\bibitem{Konoplya:2020hyk}
R.~A.~Konoplya and A.~Zhidenko,
Phys. Rev. D \textbf{101}, no.12, 124004 (2020)
doi:10.1103/PhysRevD.101.124004
[arXiv:2001.06100 [gr-qc]].

\bibitem{Rezzolla:2014mua}
  L.~Rezzolla and A.~Zhidenko,
  Phys.\ Rev.\ D {\bf 90}, no. 8, 084009 (2014)
  doi:10.1103/PhysRevD.90.084009
  arXiv:1407.3086 [gr-qc].

\bibitem{Konoplya:2016jvv}
  R.~Konoplya, L.~Rezzolla and A.~Zhidenko,
  Phys.\ Rev.\ D {\bf 93}, no. 6, 064015 (2016)
  doi:10.1103/PhysRevD.93.064015
  [arXiv:1602.02378 [gr-qc]].

\bibitem{Kofinas:2007ns}
  G.~Kofinas and R.~Olea,
  JHEP {\bf 0711}, 069 (2007)
  doi:10.1088/1126-6708/2007/11/069
  [arXiv:0708.0782 [hep-th]].

\bibitem{Takahashi:2010}
  T.~Takahashi and J.~Soda,
  Prog.\ Theor.\ Phys.\  {\bf 124}, 711 (2010)
  doi:10.1143/PTP.124.711
  [arXiv:1008.1618 [gr-qc]];
  Prog.\ Theor.\ Phys.\  {\bf 124}, 911 (2010)
  doi:10.1143/PTP.124.911
  [arXiv:1008.1385 [gr-qc]].

\bibitem{Myers:1988ze}
  R.~C.~Myers and J.~Z.~Simon,
  Phys.\ Rev.\ D {\bf 38}, 2434 (1988)
  doi:10.1103/PhysRevD.38.2434.

\bibitem{Schutz:1985km}
B.~F.~Schutz and C.~M.~Will,
Astrophys. J. \textbf{291}, L33-L36 (1985)
doi:10.1086/184453.

\bibitem{Iyer:1986np}
S.~Iyer and C.~M.~Will,
Phys. Rev. D \textbf{35}, 3621 (1987)
doi:10.1103/PhysRevD.35.3621.

\bibitem{Konoplya:2003ii}
R.~Konoplya,
Phys. Rev. D \textbf{68}, 024018 (2003)
doi:10.1103/PhysRevD.68.024018
[arXiv:gr-qc/0303052 [gr-qc]].

\bibitem{Konoplya:2019hlu}
R.~Konoplya, A.~Zhidenko and A.~Zinhailo,
Class. Quant. Grav. \textbf{36}, 155002 (2019)
doi:10.1088/1361-6382/ab2e25
[arXiv:1904.10333 [gr-qc]].

\bibitem{Konoplya:2011qq}
R.~Konoplya and A.~Zhidenko,
Rev. Mod. Phys. \textbf{83}, 793-836 (2011)
doi:10.1103/RevModPhys.83.793
[arXiv:1102.4014 [gr-qc]].

\bibitem{Gundlach:1993tp}
C.~Gundlach, R.~H.~Price and J.~Pullin,
Phys. Rev. D \textbf{49}, 883-889 (1994)
doi:10.1103/PhysRevD.49.883
[arXiv:gr-qc/9307009 [gr-qc]].

\bibitem{Bisnovatyi-Kogan:2017kii}
  G.~S.~Bisnovatyi-Kogan and O.~Y.~Tsupko,
  Universe {\bf 3}, no. 3, 57 (2017)
  doi:10.3390/universe3030057
  [arXiv:1905.06615 [gr-qc]].


\end{thebibliography}
\end{document}